\documentclass[showpacs, prd, twocolumn]{revtex4}

\usepackage{amssymb, amsmath} 
\usepackage{stmaryrd}

\usepackage{bbm} 

\usepackage[english]{babel}
\usepackage{slashed}
\usepackage{cancel}

\usepackage{graphicx}
\usepackage{color}

\usepackage{datetime}
\settimeformat{ampmtime}

\usepackage[bookmarks,breaklinks,plainpages=false,pdfpagelabels]{hyperref} 
	\hypersetup{linkcolor=red,citecolor=blue,filecolor=dullmagenta,urlcolor=darkblue} 

\definecolor{grey}{rgb}{0.4,0.4,0.4}
\definecolor{lightgrey}{rgb}{0.6,0.6,0.6}
\definecolor{dullmagenta}{rgb}{0.4,0,0.4}
\definecolor{darkblue}{rgb}{0,0,0.4}
\definecolor{orange}{rgb}{1,0.5,0}
\definecolor{lightbrown}{rgb}{0.75,0.5,0.25}
\definecolor{tan}{cmyk}{0.14,0.42,0.56,0}
\definecolor{djunglegreen}{cmyk}{0.99,0,0.52,0}
\definecolor{lightgreen}{rgb}{0,1,0}
\definecolor{olivegreen}{cmyk}{0.64,0,0.95,0.40}
\definecolor{midgreen}{rgb}{0.0,0.675,0.0}

\newcommand{\q}{\quad}
\newcommand{\qq}{\qquad}

\newcommand{\vs}{\vspace}
\newcommand{\hs}{\hspace}
\renewcommand{\.}{\hspace{0.5mm}}

\newcommand{\ra}{\ensuremath{\rightarrow}}

\newcommand{\Rrm}{\ensuremath{\mathrm{R}}}

\newcommand{\Bcal}{\ensuremath{\mathcal{B}}}

\newcommand{\Lcal}{\ensuremath{\mathcal{L}}}
\newcommand{\Mcal}{\ensuremath{\mathcal{M}}}

\newcommand{\Ocal}{\ensuremath{\mathcal{O}}}

\newcommand{\Scal}{\ensuremath{\mathcal{S}}}

\newcommand{\Abbm}{\ensuremath{\mathbbm{A}}}

\newcommand{\Xbbm}{\ensuremath{\mathbbm{X}}}

\renewcommand{\d}{\ensuremath{\mathrm{d}}}

\newcommand{\defas}{\mathrel{\mathop :}=} 
\newcommand{\hph}[1]{\hphantom{#1\;\,}}

\newcommand{\diag}{\ensuremath{\mathrm{diag}}}

\newcommand{\eg}{e.g.}
\newcommand{\ie}{i.e.}

\newcommand{\cf}{c.f.}

\makeatletter
\def\ifundefined#1{\expandafter\ifx\csname#1\endcsname\relax}
\ifundefined{default@color}
\let\default@color=\current@color
\fi
\makeatother

\newcommand{\beq}{\begin{equation}}
\newcommand{\eeq}{\end{equation}}
\newcommand{\bea}{\begin{eqnarray}}
\newcommand{\beas}{\begin{eqnarray*}}

\newcommand{\eea}{\end{eqnarray}}
\newcommand{\eeas}{\end{eqnarray*}}

\begin{document}

\title{On Instability of Certain Bi-Metric and Massive-Gravity Theories\vs{-3mm}}

\author{Florian K{\"u}hnel}
	\email{florian.kuehnel@physik.lmu.de}
	\affiliation{Arnold Sommerfeld Center, Ludwig-Maximilians University, Theresienstr.~37, 80333 M{\"u}nchen, Germany}

\date{\formatdate{\day}{\month}{\year}, \currenttime}

\begin{abstract}
\hs{-1.1mm}Stability about cosmological background solutions to the bi-metric Hassan-Rosen theory is studied.\\ The results of this analysis are presented, and it is shown that a large class of cosmological backgrounds is classically unstable. This sets serious doubts on the physical viability of the Hassan-Rosen theory{\,---\,}and in turn also of the de Rham-Gadabaze-Tolley model, to which the mentioned theory is parent. A way to overcome this instability by means of curvature-type deformations is discussed.
\end{abstract}


\maketitle

{\it Introduction \& Overview}{\;---}
The general theory of relativity proposed by Einstein in 1916 \cite{Einsteins-GR} provides the fundamental building block of our current understanding of gravitation. This framework{\,---\,}describing the dynamics of a massless spin-two field in four dimensions{\,---\,}has been tested from scales of about a fraction of a millimeter up to scales of a few astronomical units, and agrees remarkably well with all experimental data.

Despite of its successes, and the necessity of a theory of quantum gravity in the ultra-violet, it remains rather unclear whether general relativity is a valid description also on cosmological scales. Therefore it is tempting to study its consistent infra-red deformations. Several of those possibilities have been considered such as extra-dimensional models \cite{Randall:1999ee, Dvali:2000hr, Decontruction, ArkaniHamed:2002sp}, multi-gravitation \cite{Bi-Gravity, Hassan:2011zd, vonStrauss:2011mq}, or deformed (\eg~massive) gravity \cite{ArkaniHamed:2002sp, deRham:2010kj, Fierz:1939ix, Hassan:2011tf, Berkhahn:2010hc, Berkhahn:2011me}.

Since the fundamental work of Fierz and Pauli \cite{Fierz:1939ix} in 1939, who constructed a consistent theory of massive gravity on Minkowski background to {\it linear} order, the quest has long been unsuccessful to consistently generalize such a framework to curved space-times. In Ref.~\cite{Berkhahn:2011me} this task has been established on a Friedmann-Lema{\^i}tre-Robertson-Walker background, which{\,---\,}by inclusion of the Ricci scalar{\,---\,}was shown to be fully respected throughout the entire realistic cosmological evolution. An important feature of this theory is that the Fierz-Pauli mass parameter can be consistently set to zero, therefore providing a modification of general relativity solely on curved space-times. This might be very important in the light of the Boulware-Deser ghost \cite{Boulware:zf}, the v{\sc dvz} discontinuity \cite{vDVZ, On-vDVZ}, and recently-raised acausality concerns \cite{Deser-acausal}.

Many of the models which have been proposed so far to modify gravity have the unphysical need to fix a reference metric, or, if this metric is dynamical, lack the existence of a respected cosmological background. A recent and much-noticed attempt to modify gravity with a bi-metric theory which allows for cosmological backgrounds has been presented in \cite{vonStrauss:2011mq}. This work was only concerned with establishing realistic backgrounds. A complete and consistent study of fluctuations about this background is very important for stability issues (\cf~\cite{other-analyses-of-perturbations}).

In this work we present results of precisely such a stability analysis, and show that a large class of the cosmological branch of the Hassan-Rosen theory \cite{Hassan:2011zd} is not physically viable. We then show (for one particular case) a way to ensure full stability, at least on the linear level.

{\it Framework}{\;---}The bi-metric action under consideration is (\cf~Ref.~\cite{vonStrauss:2011mq})
\begin{align}
	&\Scal[ f, g, \Phi ]
		=
								- \frac{ M_{f}^{2} }{ 2 }
								\int_{\!\Mcal}\d^{4}x\;
								\sqrt{ | f |\,}\;\Rrm[ f ]
								\notag
								\displaybreak[1]
								\\[2mm]
		&
		\q						-
								\frac{ M_{g}^{2} }{ 2 }
								\int_{\!\Mcal}\d^{4}x\;
								\sqrt{ | g |\,}\;\Rrm[ g ]
								+
								\int_{\!\Mcal}\d^{4}x\;
								\sqrt{ | g |\,}\;\Lcal_{m}[ g, \Phi ]
								\notag
								\\[2mm]
		&
		\q						+
								m^{2}\.M_{g}^{2}
								\int_{\!\Mcal} \d^{4}x\;
								\sqrt{ | g |\,}\;
								\sum_{n = 0}^{4}\beta_{n}\,e_{n}
								\big(
									\Xbbm
								\big)
								\; .
								\label{eq:full-action}
\end{align}
Here, $\Xbbm \defas \sqrt{g^{-1}f}$, $\Mcal$ is a four-dimensional pseudo-Riemannian manifold, the metrics $f$ and $g$ have signature $(-, +, +, +)$, and the units are such that $\hslash \overset{!}{=} c \overset{!}{=} 1$. For the sake of convenience, the matter fields{\,---\,}which are minimally coupled to $g$ in the matter Lagrangian $\Lcal_{m}$ (we will restrict ourselves to the case of a perfect fluid){\,---\,}are denoted by $\Phi$. Hence, matter is only indirectly coupled to $f$ through its interactions with $g$. $\Rrm[ \,\cdot\, ]$ is the Ricci scalar of the respective metric, the $\beta_{n}$ are fixed, real parameters, and $e_{n}( \Xbbm )$ are elementary symmetric polynomials of the eigenvalues of the matrix $\Xbbm$, \eg~
\begin{align}
\begin{split}
	e_{0}( \Xbbm )
		&=
								1
								\; ,
								\\[2mm]
	e_{2}( \Xbbm )
		&=
								\frac{ 1 }{ 2 }
								\Big(
									\llbracket
										\Xbbm
									\rrbracket^{2}
									-
									\big\llbracket
										\Xbbm^{2}
									\big\rrbracket
								\Big)
								\, ,
\end{split}
		&
\begin{split}
	e_{1}( \Xbbm )
		&=
								\llbracket
									\Xbbm
								\rrbracket
								\; ,
								\\[2mm]
	e_{4}( \Xbbm )
		&=
								\det[ \Xbbm ]
								\; ,
\end{split}
								\label{eq:elementary-symmetric-polynominals}
\end{align}
where the double-lined square brackets denote the matrix trace, \ie~$\llbracket \Xbbm \rrbracket \equiv \Xbbm^{\mu}_{\phantom{\mu}\mu}$. The quantity $e_{3}$ is not displayed as it will not be included in the present analysis, which refers to the cosmological studies of Ref.~\cite{vonStrauss:2011mq} wherein $\beta_{3}$ is set to zero. Actually, in the case of massive gravity (where $f$ is non-dynamical) its inclusion is phenomenologically non-acceptable \cite{Koyama:2011yg}. Hence, we will set $\beta_{3} \equiv 0$.

It is easy to check that the model \eqref{eq:full-action} (except the matter sector) only depends upon three dimensionless parameters ($H_{0}$ being today's Hubble constant)\footnote{Actually, one also needs to specify how $M_{g}$ is related to the Planck mass. Here we assume that they are equal.}:
\vs{-1mm}
\begin{align}
	M_{\star}
		&\defas					\frac{ M_{f} }{ M_{g} }
								\; ,
	\q
	M
		\defas					\frac{ m }{ H_{0} }
								\; ,
	\q
	\beta_{2}
								\; .
								\label{eq:free-parameters}
\end{align}

{\it Stability}{\;---}
To check for stability or instability, respectively, one has to expand the fields $f$ and $g$ about certain backgrounds which are consistent with the action \eqref{eq:full-action}. Then one studies how the perturbations evolve.

To this end we expand $f$ and $g$ about the backgrounds $f_{(0)}$ and $g_{(0)}$, respectively,
\begin{align}
	f
		\equiv					
								f_{(0)}
								+
								\delta f
								\; ,
	\q
	g
		\equiv					
								g_{(0)}
								+
								\delta g
								\; ,
								\label{eq:define-fluctuations-about-backgrounds}
\end{align}
and define the matrix $\theta$ via $\theta^{2} \equiv g_{(0)}^{-1}\.f_{(0)}^{}$, which appears in the whole interaction term in \eqref{eq:full-action}, and allows to express $f_{(0)}$ through $g_{(0)}$ via $f_{(0)} = g_{(0)} \theta^{2}$.

In the cosmologically-relevant case, the background $g_{(0)}$ of the fluctuation $\delta g$ (to which our matter sector is coupled to) is homogeneous and isotropic, and shall assume the Friedmann-Lema{\^i}tre-Robertson-Walker form
\vs{-3mm}
\begin{subequations}
\begin{align}
	g^{}_{(0)}
		&
		=
								\diag
								\big(
									- 1,
									a^{2},
									a^{2},
									a^{2}
								\big)
								\, ,
								\label{eq:g0=diag---1--a2--a2--a2}
\end{align}
where $a$ is the scale factor, being normalized such that it equals one today. Then, demanding spatial homogeneity and isotropy for $f_{(0)}$ as well, \ie~the same ${\sf SO}( 3 )$ symmetry, and assuming the same spatial curvature as for $g_{(0)}$, leads (up to time reparametrizations) to
\vs{-1mm}
\begin{align}
	f^{}_{(0)}
		&
		=
								\diag
								\Big(\!
									- \alpha( a )^{2},
									a^{2}\.\beta( a )^{2},
									a^{2}\.\beta( a )^{2},
									a^{2}\.\beta( a )^{2}
								\Big)
								\, ,
								\label{eq:f0=diag---alpha2--a2beta2--a2beta2--a2beta2}
\end{align}
\end{subequations}
yielding
\begin{align}
	\theta
		&
		=
								\diag
								\big(
									| \alpha | ,
									| \beta | ,
									| \beta | ,
									| \beta |
								\big)
								\, .
								\label{eq:theta=diag--|alpha|--|beta|--|beta|--|beta|}
\end{align}
Hence, the functions $\alpha = \alpha( a )$ and $\beta = \beta( a )$ parametrize the deviation of the two backgrounds. In general they are not independent, as the Bianchi identity together with the conservation of energy yields
 (\cf~Ref.~\cite{vonStrauss:2011mq})
\begin{align}
	\alpha( a )
		&\equiv					\frac{ \d ( a\.\beta( a ) ) }{ \d a }
								\; .
\end{align}

The Friedmann equations determine the function $\beta( a )$. In general, it is given by a root of a quartic polynominal. However, for the choice of $\beta_{3} = 0$ (\cf~the comment on the end of the previous section), this equation is only cubic in $\beta$. Let $\rho$ be the energy density of the Universe and set $\rho_{\star} \defas \rho / 3 m^{2} M_{g}^{2}$. Then one finds
\begin{align}
\begin{split}
	\left(
		\beta_{2}
		- \frac{ 1 - \beta_{2} + 3\.M_{\star}^{2} / M^{2} }{ 3 M_{\star}^{2} }
	\right)
	\beta^{3}
	- \left(
		1
		+ 2 \beta_{2}
	\right)
	\beta^{2}
	\qq
	\\
	+ \left(
		\rho_{\star}
		+ 1
		+ \beta_{2}
		- \frac{ \beta_{2} }{ M_{\star}^{2} }
	\right)
	\beta
	+ \frac{ 1 + 2 \beta_{2} }{ 3 M_{\star}^{2} }
		=
								0
								\; .
								\label{eq:cubic-beta-equation}
\end{split}
\end{align}

\begin{figure}[t]
	\vs{-3mm}
	\begin{center}
		\includegraphics[scale=0.40]{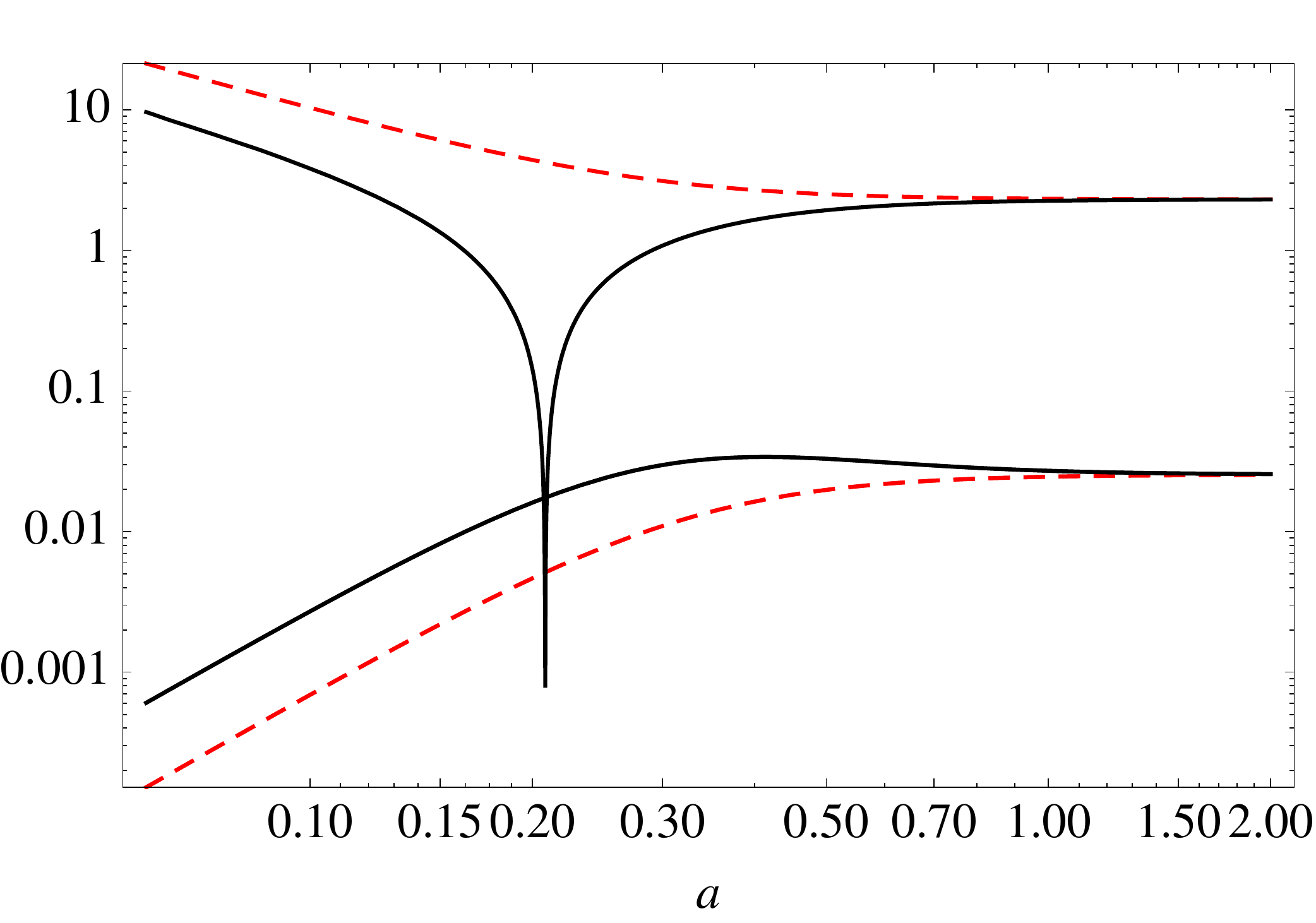}
		\caption{
				Absolute values of the functions $\alpha( a )$ (black, solid curves) 
				and $\beta( a )$ (red, dashed curves)
				as functions of the scale factor $a$ (double-logarithmic scale). 
				The parameters are $\beta_{2} = -0.3$, $M = 3$, $M_{\star} = 2.5$.
				Note that the two lower curves are two-fold degenerate.\vs{-3.5mm}
				}
		\label{fig:betasol}
	\end{center}
\end{figure}

The corresponding three solutions to Eq.~\eqref{eq:cubic-beta-equation} show a quite different behavior (\cf~Fig.~\ref{fig:betasol}). In fact, in the limit of small scale factor one of them diverges, while two approach zero. The latter solutions implie that those terms (involving powers of $\delta f_{\mu \nu}$), that are contracted with the inverse metric $f_{(0)}^{-1}$, have prefactors that strongly grow (and eventually diverge in the limit of vanishing scale factor) the stronger the higher their order is.

Actually, as we will see below, fluctuations become of order one already at some moderately small value of $a$. Moreover, one of the solutions has a zero-crossing of $\alpha( a )$ (\cf~Fig.~\ref{fig:betasol}), which makes $\theta( a )$ {\it non-analytic}{\,---\,}a particularity which also concerns the solutions for the fluctuations (see below).

Let us now come to the general cosmological case, as discussed in Ref.~\cite{vonStrauss:2011mq}. Assuming a spatially-flat Universe, one can show that the Friedmann equation takes the form $( H / H_{0} )^{2} = \Omega + \Omega_{\beta} \equiv ( \Omega_{\text{r}} + \Omega_{\text{m}} + \Omega_{\Lambda}) + \Omega_{\beta}$, wherein $\Omega_{\text{r}, \text{m}, \Lambda}$ denote the density parameters for radiation, matter, and a cosmological constant, respectively. As usual we define $H \defas \dot{a} / a$, $\Omega \defas \rho / 3 H_{0}^{2} M_{g}^{2}$, and further set $\Omega_{\beta} \defas M^{2} ( \beta - 1 ) [ \beta_{2} ( \beta + 1 ) - ( 1 + 2\.\beta_{2} )\beta_{2} ]$. We demand that $\beta( a = 1 ) = 1$ in order to have that the density parameter $\Omega$ equals one today, \ie~$\Omega( a = 1 ) = 1$, being suggested by {\sc cmb} observations \cite{Ade:2013zuv}.

By performing distance-related tests using cosmological data, it has been shown by the authors of Ref.~\cite{vonStrauss:2011mq} that it is possible to choose the above parameters such that a realistic cosmological background can be obtained. Unfortunately, this has been done only for a very limited range of red shifts, and is purely on the background level.

For the sake of studying stability, the standard way is to perform a decomposition of the fluctuations $\delta g$ and $\delta f$ into irreducible tensors with respect to the isometries of the Friedmann backgrounds. In this way, and on the linear level, one can study the rank-2,1,0 ${\sf SO}( 3 )$-tensor contributions separately. Often, the scalar sector is the most (in)stability-indicative one. Precisely the same results can, however, be obtained in the following way: As we are interested in studying stability of the homogeneous backgrounds, it suffices to look at the fluctuations' zero modes. Then, it is easy to show that the metrics' off-diagonal spatial components ($i \ne j$) can be solved for separately, and are invariant with respect to time-reparametrizations.

After expanding the action \eqref{eq:full-action} to second order in the fluctuations, we find for $i \ne j$ the set of coupled field equations
\begin{subequations}
\begin{align}
	\delta f_{ij}''
	+
	a_{1}\.\delta f_{ij}'
	+
	b_{1}\.\delta f_{ij}
		&=
								c_{1}\.\delta g_{ij}
								\label{eq:EOM-ij-a}
								\; ,
								\displaybreak[1]
								\\[2mm]
	\delta g_{ij}''
	+
	a_{2}\.\delta g_{ij}'
	+
	b_{2}\.\delta g_{ij}
		&=
								c_{2}\.\delta f_{ij}
								\; ,
\end{align}
\end{subequations}
wherein a prime denotes a derivative with regard to the scale factor $a$, and the quantities $a_{i}$, $b_{i}$, $c_{i}$ are given by
\begin{align}
	a_{1}
		&=
								- \log{'}\!\left[ a\.\tau\.\alpha\.\beta \right]
								\, ,
								\notag
								\displaybreak[1]
								\\[1.5mm]
	\q
	a_{2}
		&=
								- \log{'}\!\left[ a\.\tau\.\right]
								\, ,
								\notag
								\displaybreak[1]
								\\[2mm]
	b_{1}
		&=
								\frac{M^{2}\.\tau^{2}}
								{M_{\star}^{2}}
								\Bigg[
									\beta_{2}\.|\alpha|^{3}\.| \beta |
									+
									2\.|\alpha|^{3}\.| \beta |^{3}
									\bigg(
										1
										+
										3\.\frac{ M_{\star}^{2} }{ m^{2} }
										-
										\beta_{2}
									\bigg)
								\notag
								\displaybreak[1]
								\\[0mm]
		&
		\hphantom{=\;}
									+
									\alpha^{2}\.| \beta |
									\Big[
									\beta_{2}
									\big(
										3\.| \beta |
										-
										2
									\big)
									-
									1
									\Big]
									+ \tfrac{ 4\.M_{\star}^{2} }{ M^{2} a^{2} \tau^{2} \beta^{2}}\!
									\left(a \beta \right)'^{2}
								\Bigg]
								\. ,
								\\[2mm]
	b_{2}
		&=
								M^{2}\.\tau^{2}
								\Big[
									6
									+
									3
									| \beta |
									\big[
										\beta_{2}
										(
											| \alpha |
											-
											2
										)
										-
										1
									\big]
									-
									\tfrac{ 2\.\Omega_{r}}{ M^{2}\.a^{4} }
									+
									\beta_{2}\.\beta^{2}
								\notag
								\displaybreak[1]
								\\[0mm]
		&\hph{=}
								\phantom{M^{2}\.\tau^{2}\Big[}
									+
									\tfrac{ 4 }{ a^{2}\.M^{2}\tau^{2} }
									+
									\tfrac{ 6\.\Omega_{\Lambda} }{ M^{2} }
									-
									2
									(
										1
										+
										2\.\beta_{2}
									)
									| \beta |
									+
									6\.\beta_{2}
								\Big]
								\. ,
								\notag
								\displaybreak[1]
								\\[2mm]
	c_{1}
		&=
								-\,
								\frac{M^{2}\.\tau^{2}\.\alpha^{2}\.|\beta|^{3}}
								{M_{\star}}
								\Big[
									\beta_{2}
									\left(
										|\alpha|
										+
										|\beta|
										- 2
									\right)
									-
									1
								\Big]
								\. ,
								\notag
								\displaybreak[1]
								\\[2mm]
	c_{2}
		&=
								c_{1}\.\alpha^{-2}\.\beta^{-4}
								\; ,
								\notag
\end{align}
with $t$ being cosmic time, and $\tau \defas H_{0}\.\d\mspace{1mu}t( a ) / \d a$. Defining
\begin{align}
\begin{split}
	y^{(1)}_{ij}
		&\defas					\frac{ \delta f_{ij} }{ f^{(0)}_{ii} }
								\; ,
	\q
	y^{(2)}_{ij}
		\defas					\left( \frac{ \delta f_{ij} }{ f^{(0)}_{ii} } \right)^{\!\!'}
								,
								\\[1mm]
	y^{(3)}_{ij}
		&\defas					\frac{ \delta g_{ij} }{ g^{(0)}_{ii} }
								\; ,
	\q
	y^{(4)}_{ij}
		\defas					\left( \frac{ \delta g_{ij} }{ g^{(0)}_{ii} } \right)^{\!\!'}
								,
								\label{eq:}
\end{split}
\end{align}
one can express the system (\ref{eq:EOM-ij-a},b) as
\begin{align}
	\vec{y}\.'
		&=
								\Abbm \cdot \vec{y}
								\; ,
								\label{eq:system}
\end{align}
wherin the matrix $\Abbm$ is composed of the coefficients $a_{i}$, $b_{i}$, $c_{i}$. Stability of the above system \eqref{eq:system} depends upon the behavior of the real parts of the eigenvalues of $\Abbm$.

\begin{figure}[t]
	\vs{-3mm}
	\begin{center}
		\includegraphics[scale=0.40]{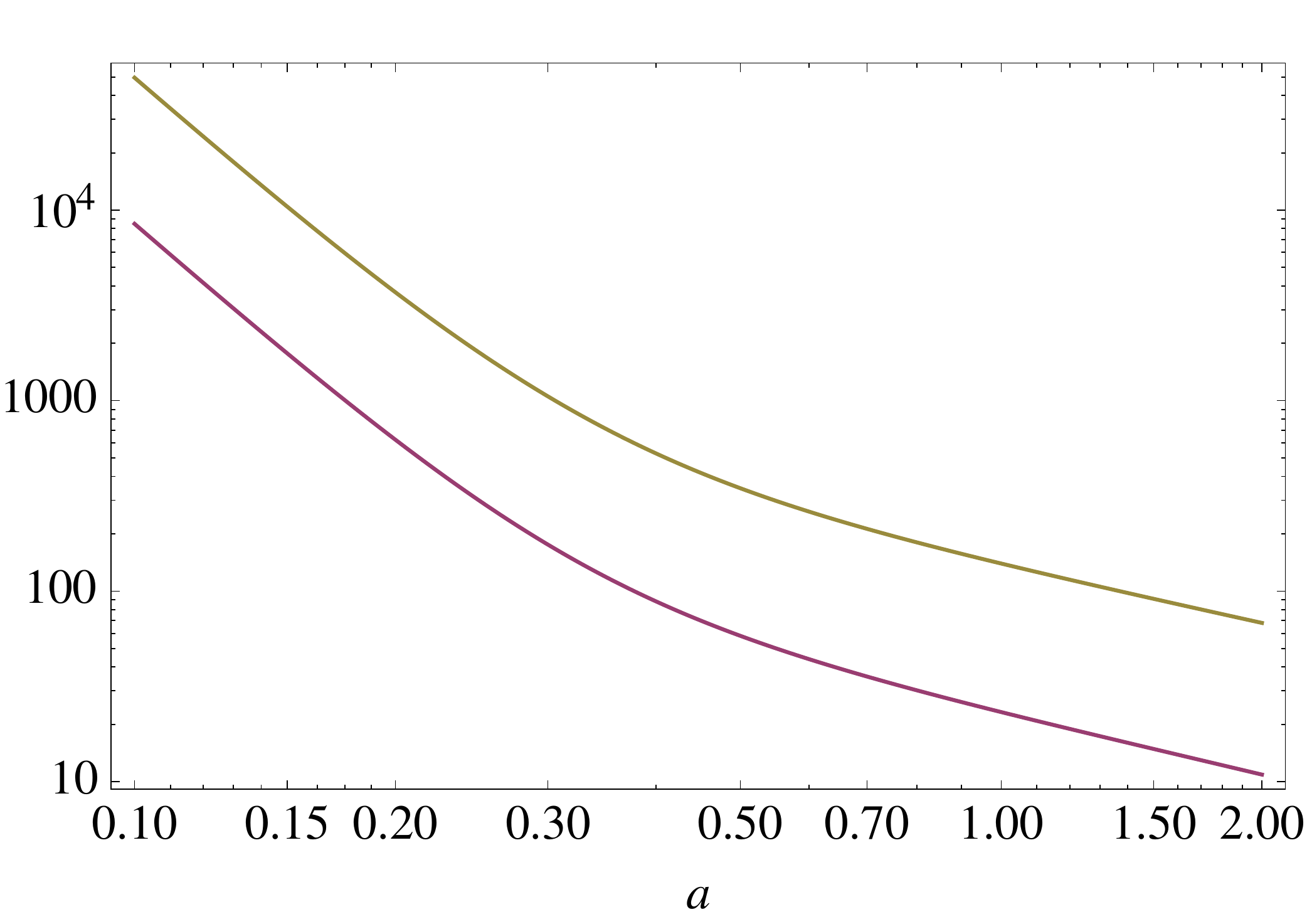}
		\includegraphics[scale=0.40]{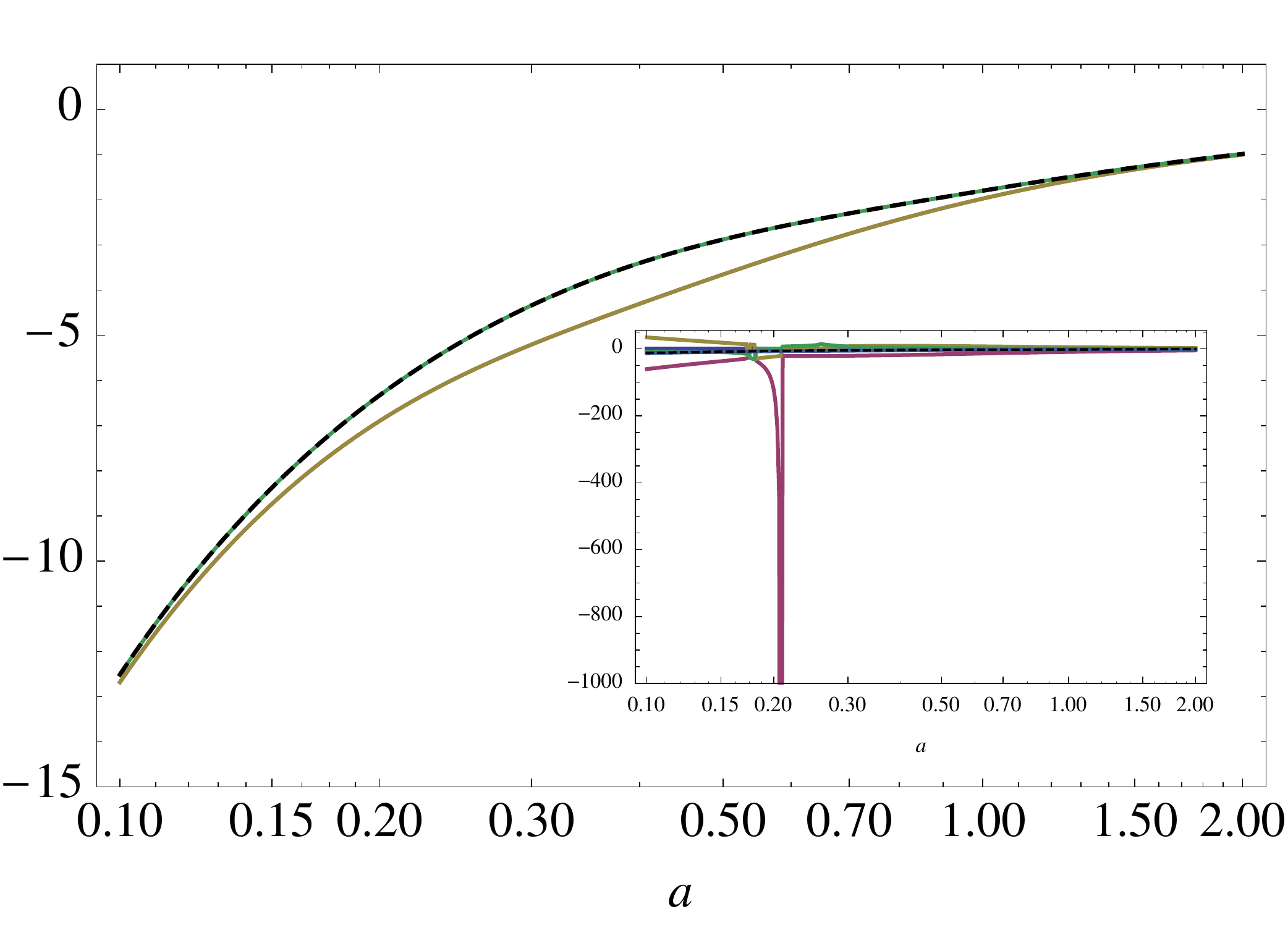}
		\caption{
				Real parts of the eigenvalues of the matrix $\Abbm$ (\cf~Eq.~\eqref{eq:system})
				as functions of the scale factor $a$ (log-axis). The two large panels represent the 
				regular solutions to Eq.~\eqref{eq:cubic-beta-equation}, and the small graph in 
				the lower panel shows the one for which $\alpha( a )$ is non-analytic. Dashed lines 
				indicate the eigenvalues of the undeformed theory. 
				The parameters are $\beta_{2} = -0.3$, $M = 3$, $M_{\star} = 2.5$.
				}
		\label{fig:EV}
	\end{center}
\end{figure}

Analyzing (numerically) precisely those real parts shows (\cf~Fig.~\ref{fig:EV}), first of all, that the undeformed theory (\ie~zero interactions) is stable (dashed lines in the lower panel of Fig.~\ref{fig:EV}). For the deformed theory one observes that there is always (\ie~for all three solutions to Eq.~\eqref{eq:cubic-beta-equation}) at least one eigenvalues which diverge towards $+ \infty$ if $a$ goes to zero. More precisely, for $a \ra 0$ it diverges much faster than $1 / a$, yielding an exponential divergence of the associated mode. Furthermore, the solution to Eq.~\eqref{eq:cubic-beta-equation} for which $\alpha( a )$ and $\beta( a )$ grow for small $a$ (and hence does {\it not} imply that higher-order terms in the expansion of the action \eqref{eq:full-action} become more and more important as $a$ becomes smaller (\cf~remark on the previous page)) and which has a zero-crossing, diverges at some finite value $a = a_{\star}$. The divergence is such that it grows towards $- \infty$ for $( a_{\star} - a ) \ra 0^{+}$.

\begin{figure}[t]
	\begin{center}
		\vs{-2mm}
		\includegraphics[scale=1.0]{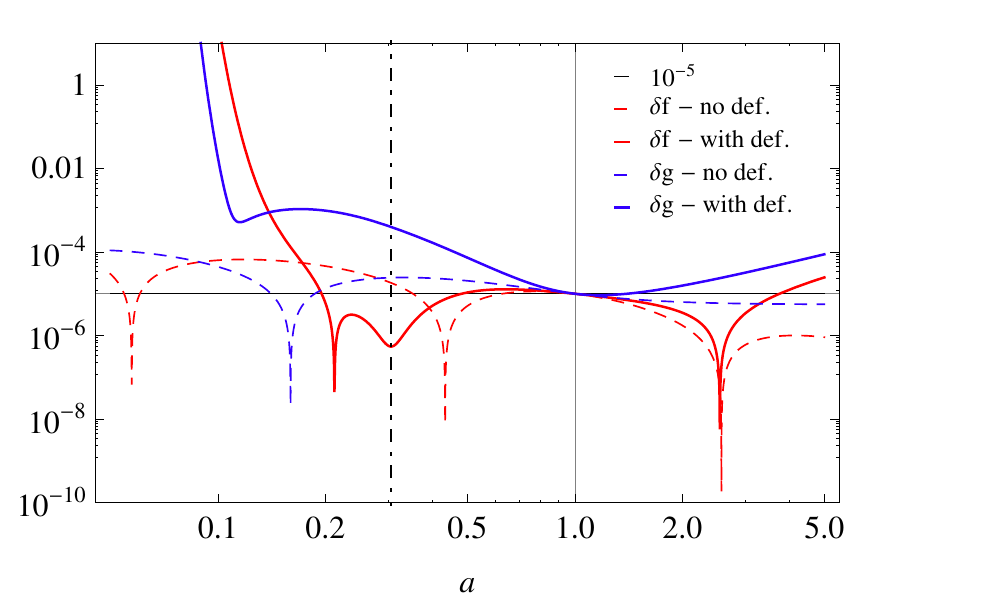}
		\caption{
				Absolute values of the relative metric fluctuations 
				$y^{(1)}_{23} \defas \delta f^{}_{23} / f^{(0)}_{22}$ 
				(red curve) and $y^{(3)}_{23} \defas \delta g^{}_{23} 
				/ g^{(0)}_{22}$ (blue curve) as functions of the 
				scale factor $a$. Dashed, colored lines correspond 
				to the undeformed theory, solid lines to the 
				deformed one. The dot-dashed, vertical line is at 
				$a = a_{\star} \approx 0.3$ (for the parameter set 
				\eqref{eq:parameter-set}).}
		\label{fig:solution-to-EOM}
	\end{center}
\end{figure}

We should stress that all parameters within the physi-cally-relevant intervals (as given in Ref.~\cite{vonStrauss:2011mq})
\vs{-1mm}
\begin{align}
\begin{split}
	1.5
		\lesssim
								M_{\star}
								\lesssim
								3.0
								\, ,
	\;
	2
		\lesssim
								M
								\lesssim
								3.5
								\, ,
								\\[1.5mm]
	-0.5
		\lesssim
								\beta_{2}
								\lesssim
								- 0.1
								\, ,
								\mspace{46mu}
								\label{eq:parameter-intervals}
\end{split}
								\\[-7mm]
								\notag
\end{align}
yield the same qualitative behavior. In all those cases the scale factor at which the theory is non-analytic, $a_{\star}$, is far larger than its value at recombination. On the other hand, it is smaller than the value up to which super-nov{\ae} data have been analyzed in \cite{vonStrauss:2011mq}. Exactly the same holds true for the value at which fluctuations become of order one, $a_{\text{nl}}$ (see below). Choosing
\vs{-1mm}
\begin{align}
	M_{\star}
		&=						2.5
								\; ,
	\q
	M
		=						3.0
								\; ,
	\q
	\beta_{2}
		=						- 0.3
								\; ,
								\label{eq:parameter-set}
\end{align}
we find that both $a_{\text{nl}}$ and $a_{\star}$ are $\Ocal( 0.1 )$\! (\cf~Figs.~\ref{fig:EV} \& \ref{fig:solution-to-EOM}).

\begin{figure}[t]
	\begin{center}
		\vs{-5mm}
		\includegraphics[scale=0.03]{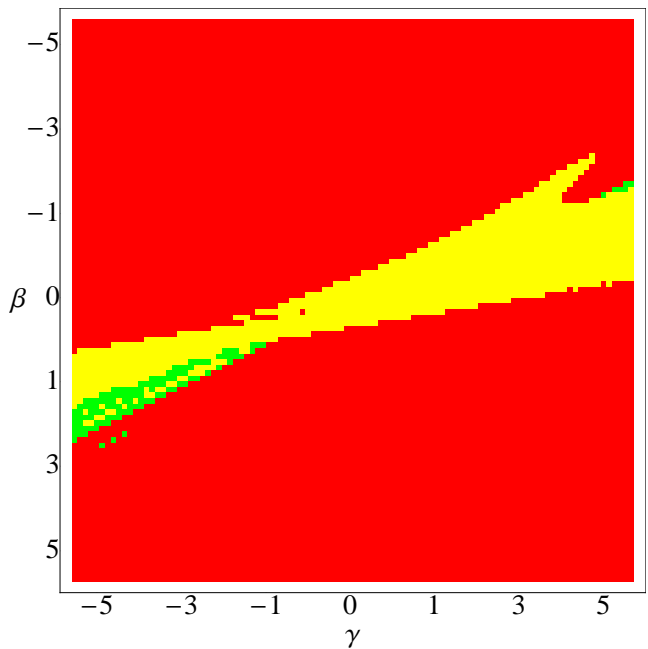}
		\caption{
				Stability parameter plot in the $\beta$-$\gamma$ plane,
				exemplary for $\alpha = 0$, $\beta_{2} = - 0.4$ and $M = 1$.
				Green dots represent fully stable regions,
				yellow ones indicate classical instability,
				and red points stand for unitarity violation (\cf~main text).\vs{-5mm}}
		\label{fig:gamma-beta-parameter-plot}
	\end{center}
\end{figure}

Fig.~\ref{fig:solution-to-EOM} shows the full solution to the system (\ref{eq:EOM-ij-a},b). It can be seen (exemplary for the parameter set \eqref{eq:parameter-set}, and certain initial conditions) how the (absolute values of the) relative fluctuations behave as a function of the scale factor $a$. One can read off the aforementioned instability from the solid lines, describing one particular realization of the deformed theory. In contrast, the undeformed theory (dashed lines) is well-behaved.

One also observes the same unphysical {\it backward} instability as in Refs.~\cite{Berkhahn:2010hc}, implying that the set-up is merely self-protected, where the notion of 'self-protection' refers to the breakdown of the linear approximation, \ie, the formation of a new background, such that no unitarity violation can be seen within this approximation.

We checked that the instability occurs for all cosmologically-allowed parameters out of the intervals \eqref{eq:parameter-intervals} (for the present case of no matter coupling to $f$, as discussed in Ref.~\cite{vonStrauss:2011mq}). On top of that, it is independent of the precise details of the initial conditions.\footnote{As mentioned earlier, we could have equivalently performed our analysis in another language, \eg, that of Ref.~\cite{Berg}. In that notation one finds precisely the same divergence in the {\it gauge-invariant} quantity $\Bcal$, being composed out of parts of the fluctuations' off-diagonal parts.}

Let us finally come to one particularly interesting case, which is constituted by the limit $M_{\star} = M_{f} / M_{g} \ra \infty$. Therefor the $f$-field is frozen into its background value, which may be taken to be Minkowskian due to the lack of respective matter couplings (\cf~Eq.~\eqref{eq:full-action}). Performing analogous studies as above reveals the same mentioned backward instability{\,---\,}the figure corresponding to Fig.~\ref{fig:solution-to-EOM} looks qualitatively the same in this respect. Since, now, there is only one dynamical metric (albeit with a particular deformation term) we can easily use a modified version of the stability analysis performed in Ref.~\cite{Berkhahn:2011me}. This tantamounts to study{\,---\,}after introduction of St{\"u}ckelberg fields{\,---\,}the roots of the determinant of the full kinetic operator, from which bounds for stability and unitarity can be directly read off.

Explicitly, and following Ref.~\cite{Berkhahn:2011me}, one introduces St{\"u}ckelberg fields as
\vs{-1mm}
\begin{align}
	\delta g_{\mu\nu}
		&=						h_{\mu\nu} 
								+ \nabla_{\!(\mu} A_{\nu)}
								+ \nabla_{\!\mu}\nabla_{\!\nu} \Phi \;.
								\label{eq:H}
\end{align}
Here, $h$, $A$, $\Phi$ are rank-2,1,0 tensors, respectively, under full background diffeomorphisms; round brackets around indices stand for symmetrization. This parametrization corresponds to two successive St{\"u}ckelberg completions and introduces a $U(1)^4\times U(1)$ gauge symmetry among the fields $h$, $A$, $\Phi$.

The task is now to supplement the linearized action \eqref{eq:full-action} with a 'healthy' deformation term, such that the theory respects realistic cosmological backgrounds, \ie~those FLRW ones as in Eq.~\eqref{eq:g0=diag---1--a2--a2--a2} which are in agreement with observations.

The Goldstone-St{\"u}ckelberg field $\Phi$ enters the gauge-invariant combination $\delta g$ with two derivatives and, therefore, {\it a priori} any modified quadric action with four derivatives. Without further restriction, the short distance behavior of the deformation would be governed by a higher-derivative theory that violates unitarity. In order to avoid pathological four-derivative terms, and to second adiabatic order (given by the number of derivatives acting on the background metric) one can show that the unique way of proper covariantization is given by adding to the Lagrangian curvature-type deformations of the form
\vs{-1mm}
\begin{align}
\begin{split}
		&						\delta g_{\mu \nu}
								\bigg[
								\alpha\.R^{}_{0}\.
								g_{0}^{\; \mu[ \nu}\.g_{0}^{\; \beta]\alpha}
								+
								\gamma\.R_{0}^{\;\mu\alpha\nu\beta}
								\\[0.0mm]
		&\phantom{\delta g_{\mu \nu}\bigg[}			
								+
								\beta
								\Big(\.
								R_{0}^{\;\mu [\nu}\.g_{0}^{\;\beta]\alpha}
								+\.R_{0}^{\; \alpha[ \beta}\.g_{0}^{\;\nu] \mu}
								\Big)
								\bigg]
								\delta g_{\alpha \beta}
								\;.
								\label{eq:M_supplement}
\end{split}
\end{align}
Here, a subscript $0$ indicates a $g_{0}$-background quantity, $\alpha, \beta, \gamma$ are real dimensionless parameters, and square brackets around indices stand for anti-symmetrization. Including terms of higher adiabatic order requires introducing further parameters with appropriate inverse mass dimension to compensate for the additional derivatives acting on $g_{0}$.

The stability analysis only requires to determine the roots of the determinant of the kinetic operator of the new action, which signal the saturation of the stability or unitarity bounds \cite{Berkhahn:2010hc}, respectively. Crossing the first bound indicates the breakdown of the linear approximation and the formation of a new background. Crossing the latter indicates an inconsistency, as it means that the system looses its probabilistic interpretation.

In order to calculate the determinant of that kinetic operator it is useful to completely fix the gauge to $h_{0 \mu} = 0$ and $A_{0} = 0$. Then, unitary violation is indicated by the zero crossing of the coefficient in front of the highest power in the temporal component of the momentum. Classical stability is determined by the zero crossing of the coefficient in front of the highest power in the spatial components of the momentum. 

Fig.~\ref{fig:gamma-beta-parameter-plot} shows a respective (in)stability parameter plot ($\beta$-$\gamma$ plane), exemplary for $\alpha = 0$, $\beta_{2} = - 0.4$ and $M = 1$. One observes, in particular, two things: First, the case of $\alpha = \beta = \gamma = 0$ (which corresponds to the original model) is classically unstable, albeit it does not violate unitarity (as expected), and provides an independent confirmation of the aforementioned instability. Second, there exists parameter values (being of order one) such that the linear theory is truly stable.

However, this {\it necessarily} involves curvature extensions (\cf~Eq.~\eqref{eq:M_supplement}). For the full model in which both metrics (spin-two fields) are dynamical, the situation seems problematic due to inevitable kinematic modifications. We will devote a future publication to such an analysis.\\

{\it Summary \& Outlook}{\;---}
Let us summarize: Starting from the general bi-metric model \eqref{eq:full-action}, using the general phenomenologically-viable parameters intervals \eqref{eq:parameter-intervals} (given in Ref.~\cite{vonStrauss:2011mq}) that allow for cosmological backgrounds, and expanding the action to second order in the fluctuations about these backgrounds, one finds that the theory under consideration is classically unstable.

We confirmed (for a special case) our results with an independent analysis method (introduced in Ref.~\cite{Berkhahn:2011me}), and were able to show{\,---\,}by appropriate supplementation with the curvature-type deformation terms \eqref{eq:M_supplement}{\,---\,}that the theory (with one metric being frozen) can be made stable on the linear level.

The full non-linear bi-metric theory{\.---\.}which is background independent{\.---\.}might, however, not allow to cure the aforementioned instabilities in the described way. This is so because the curvature terms will, then, be applied on dynamical metric(s) and not only on the background metric(s). Terms like, \eg, $\Rrm_{f}[ f ]$, $\Rrm_{g}[ g ]$, $\Rrm_{g}[ f ]$, $\Rrm_{f}[ g ]$, \ldots, must occur in front of the potential term in order to generate the mentioned background curvature terms. This necessarily involves kinetic modifications, also in the tensor sector, in such a way that ghosts are difficult, if not impossible, to avoid. So, also in the light of recent acausality concerns \cite{Deser-acausal}, it might very well be that nature prefers undeformed and massless gravity.

{\it Acknowledgments}{\;---}
It is a pleasure to thank Lasma Alberte, Felix Berkhahn, Gia Dvali, Cristiano Germani, Stefan Hofmann, Michael Kopp, Florian Niedermann, Angnis Schmidt-May and Robert Schneider for helpful discussions. Furthermore, I would like to thank the anonymous referee for valuable comments regarding presentation and discussion. This work was supported by the Excellence Cluster 'Origin and Structure of the Universe'.\\[5mm]


\end{document}